\RequirePackage{fix-cm}
\documentclass[smallextended]{svjour3}       % onecolumn (second format) for METRIKA
\smartqed  % flush right qed marks, e.g. at end of proof
\usepackage{graphicx}
\usepackage{multirow}
\usepackage{rotating}
\usepackage{color}

\newcommand{\eg}{\end{equation}}
\newcommand{\bg}{\begin{equation}}
\definecolor{darkred}{rgb}{0.9,0.1,0.1}
\newcommand{\new}[1]{\textcolor{darkred}{#1}} % new text
\newcommand{\old}[1]{} % remove text
\journalname{Metrika}

\begin{document}

\title{Cuboidal Dice and Gibbs Distributions}

\author{Wolfgang Riemer \and Dietrich Stoyan \and Danail Obreschkow}

\institute{Wolfgang Riemer \at
              August-Bebel-Str. 80, Pulheim 50259, Germany \\
              \email{w.riemer@arcor.de}\\
           \and
           Dietrich Stoyan \at
           Institut f\"{u}r Stochastik, TU Bergakademie Freiberg, Freiberg 09596, Germany \\
           \email{stoyan@math.tu-freiberg.de} \\
           \and
           Danail Obreschkow \at
           International Centre for Radio Astronomy Research, M468, University of Western Australia \\
           35 Stirling Hwy, Crawley, WA 6009, Australia\\
           \email{danail.obreschkow@uwa.edu.au}
}

\date{Received: 11/02/2013 / Accepted: 15/02/2013} % The correct dates will be entered by the editor

\maketitle

\noindent\new{In this version, typographical mistakes made in the original paper (published in Metrika, 2013), have been  corrected. All corrections are highlighted in red.}\\

\begin{abstract}
What are the face-probabilities of a cuboidal die, i.e.~a die with different side-lengths? This paper introduces a model for these probabilities based on a Gibbs distribution. Experimental data produced in this work and drawn from the literature support the Gibbs model. The experiments also 
reveal that the physical conditions, such as the quality of the surface onto which the dice are dropped, can affect the face-probabilities. In the Gibbs model, those variations are condensed in a single parameter, adjustable to the physical conditions.
% 01.50.Wg = Physics of Toys
% 02.50.Cw = Probability Theory
% 02.50.Fz = Stochastic analysis
% 05.20.Gg = Classical ensemble theory
% 05.45.Gg = Control of chaos, applications of chaos\
\end{abstract}

\section{Introduction}
Symmetry considerations make it seem obvious that an ideal cubical die lands on all six faces with an identical probability of $\frac{1}{6}$. However, what happens when non-fair
dice are tossed? In particular, what are the face-probabilities of a homogeneous \emph{cuboid}, i.e.\ a six-sided polyhedron with parallel faces but different side-lengths? This is a surprisingly challenging problem.

This paper offers a robust answer to the question above. It begins with a brief historical account (Section \ref{section_history}) and presents a control experiment with a single cuboid (Section \ref{section_experiment}) that is used as a benchmark for the theoretical modelling. Section \ref{section_gibbs} then introduces a new model based on a Gibbs distribution, which is found to be consistent with the control experiment. This model naturally contains a free parameter, which characterizes the physical conditions of the experiment. The new model is then compared against experimental data with differently sized cuboids drawn from the literature (Section \ref{section_classical}) and extended to non-cuboidal dice via the example of U-shaped dice (Section \ref{section_extension}). Section \ref{section_summary} summaries the findings of this paper. 

\section{Brief history}\label{section_history}

Already Isaac Newton mentioned the problem. In a private writing dated between 1664 and 1666, and
published in 1967 (Newton, 1967, p.~60--61), he wrote on the face-probabilities of a tossed cuboid: ``if a die bee not a Regular
body but a Parallelepipedon or otherwise unequally sided, it may bee
found how much one cast is more easily gotten then another.'' It remains unclear whether Newton really tried to
solve this problem.

In 1692, the problem appeared again in a paper by John Arbuthnot: ``In a Parallelopipedon, whose Sides are to
another in the Ratio of $a$,$b$,$c$: to find at how many Throws any
one may undertake that any given Plane, viz. $ab$, may arise'' (quoted from Hyk\v{s}ov\'a et al., 2012). Arbuthnot wrote that he
left ``the solution to those who think it merits their pains.''

Fifty years later, Thomas Simpson (1740) used a simple geometrical idea to model the face-probabilities of a tossed cuboid. He
assumed the probability of each face to be proportional to the surface area of the
corresponding spherical quadrilateral, i.e.\ to the solid angle spanned by the face when seen from the centre of the cuboid.
However, subsequent experimental investigations (e.g.~Singmaster, 1981) clearly rejected Simpson's model. Budden (1980) and Heilbronner (1985) also experimented with
series of cuboids. Although Budden and Heilbronner did not find a formula for the face-probabilities, their data again disqualifies the Simpson model, and so do modern computer simulations of
tossed cuboids (Obreschkow, 2006). Regardless of the clear insufficiency of the Simpson model, a recent paper by Hyk\v{s}ov\'a et al.~(2012) still refers to this model without criticism. 
Because of this discrepancy, this paper will first reemphasize the insufficiency of the Simpson model (Section \ref{section_experiment}), before introducing a much more accurate model (Section \ref{section_gibbs}).

\section{Control experiment}\label{section_experiment}

The control experiment is performed with a wooden ($13\times20\times23\rm~mm^3$)-cuboid. When tossing this cuboid, it became clear that the face-probabilities significantly depend on the physical conditions, such as the tossing
technique, the height of free fall, the shape of the cuboid's edges, and the elasticity of the surface on which the cuboid lands. For example, a rough or elastic surface generally increases the face-probabilities of the two largest faces. The same qualitative
change is observed when the cuboid is tossed from an arm-length above the table rather than using a dice cup.

To account for the importance of the physical conditions, two experimental runs were performed. In experiment I, the cuboid was tossed $N=2,700$ times on a wooden table using a leather dice cup. In experiment II, the cuboid was dropped $N=1,000$ times onto a polished steel surface from an initial height of 1~m. Table \ref{table_control} lists the observed frequencies $f_i=n_i/N$, where $n_i$ is the number of times that face $i$ $(i=1,...,6)$ showed up. As expected, the measured frequencies differ significantly between experiment I and II, thus demonstrating that the shape of the cuboid alone does not determine the face-probabilities. Several physical reasons might be responsible for the different outcome probabilities in the two experiments. For example, the dice cup might have a stabilizing function when the cuboid lands on one of its small faces. In turn, dropping a cuboid from a high level (1~m) implies that the cuboid bounces off the floor many times before it comes to rest. Multiple bounces tend to result in a fast rolling around the longest axis of the cuboid, which implies that the cuboid is very unlikely to land on one of the two smallest faces. Qualitatively this suggests that the face-probabilities of the largest faces increase with the initial energy of the tossing process.

The differences in the observed frequencies of opposite faces (e.g.~face 3 and 4) give a rough estimate of the deviation between measured frequencies and underlying probabilities. Those deviations would disappear if $N\rightarrow\infty$.

For comparison Table \ref{table_control} shows the face-probabilities predicted by the Simpson model (explained in Section \ref{section_history}). This model fits neither of the two experiments. In comparison to both experiments, it clearly overpredicts the probabilities of the smallest faces and underestimates the probabilities of the largest faces. A much more accurate description is offered by the Gibbs models in Table \ref{table_control}, whose free parameter has been fitted to experiment I and II, respectively. This new model is now explained in Section \ref{section_gibbs}.

\begin{table}[t]
	\centering
	\begin{tabular}{|l|c|c|c|c|c|c|}
		\hline
		\bf{Face $i$} & \bf{1} & \bf{2} & \bf{3} & \bf{4} & \bf{5} & \bf{6} \\
		\hline
		Surface area [mm$^2$] & 299 & 260 & 460 & 460 & 260 & 299 \\
		Half-height $h_i$ [mm] & 10 & 11.5 & 6.5 & 6.5 & 11.5 & 10 \\
		\hline
		$f_{i}$ experiment I ($N=2,700$) [\%] & 10.3 & 7.7 & 30.9 & 32.7 & 7.6 & 10.9\\
		$f_{i}$ experiment II ($N=1,000$) [\%] & 5.5 & 1.5 & 43.5 & 42.5 & 2.6 & 4.1\\
		\hline
		$p_{i}$ Simpson model [\%] & 13.5 & 10.5 & 26.0 & 26.0 & 10.5 & 13.5\\
		$p_{i}$ Gibbs model ($\beta=\new{2.69}\old{4.90}$) [\%] & 11.2 & 7.2 & 31.6 & 31.6 & 7.2 & 11.2\\
		$p_{i}$ Gibbs model ($\beta=\new{5.58}\old{10.2}$) [\%] & 5.0 & 2.0 & 43.0 & 43.0 & 2.0 & 5.0 \\
		\hline
	\end{tabular}
	\caption{Control experiment with a homogeneous ($13\times20\times23\rm~mm^3$)-cuboid. Faces 1 and 6: $13\times23$~mm; faces 2 and 5: $13\times20$~mm, faces 3 and 4: $20\times23$~mm. See Section \ref{section_experiment} for details.}
	\label{table_control}
\end{table}

\section{Gibbs distribution}\label{section_gibbs}

Following independent ideas of Riemer (1991) and Obreschkow (2006), this section uses Gibbs distributions
to model the face-probabilities of tossed cuboids. Gibbs distributions are probability distributions that are commonly used
in many fields of probability theory, mathematical statistics, as well as statistical mechanics, from where they
originate. The philosophy of this paper is to adopt Gibbs distributions in a heuristic way, that is without
deriving them from a set physical assumptions. The model then gains its validity \emph{a posteriori} though verification
against experimental data -- a common approach in statistics.

A Gibbs distribution can be summarized as follows: consider a system
with $k$ states, where each state $i=1,...,k$ has a positive energy
$E_i$. If the Gibbs theory applies, the system is found in state $i$ with probability
\bg \label{Gib}
p_i(\beta) = Z(\beta)^{-1} \exp(-\beta E_i),
\eg
where $\beta$ is a positive parameter, called inverse temperature (because it is proportional to $T^{-1}$ in thermodynamics), and $Z(\beta)\equiv\sum_i\exp(-\beta E_i)$ is a normalization factor, called the partition function. The parameter $\beta$ controls the character of the Gibbs
distribution: if $\beta=0$ the distribution is
uniform with equal probabilities for all states $i\in\{1,...,k\}$; as $\beta\rightarrow \infty$ the distribution becomes peaked with the minimal energy state(s) having a probability equal to 1; for any intermediate $\beta\in(0,\infty)$, the probability of a state increases monotonically with decreasing energy.

In modeling the tossing experiments, the states are the faces that end up lying on top, i.e.~the cuboid is said to be in state $i$ if it comes to rest with face $i$ on top (thus $k=6$). The energy of
state $i$ is taken proportional to the potential energy, i.e.~to the height $h_i$ of the center of gravity in state $i$. Note that in this way inhomogeneities in the mass distribution of the cuboid are accounted for, as illustrated in Section \ref{section_extension}. If the cuboid is homogeneous,
$h_i=s_i/2$ where $s_i$ is the vertical side-length of the cuboid in state $i$.
To eliminate physical units, the energy $E_i\propto h_i$ is normalized to \new{$h_0=0.5\,volume^{1/3}$}\old{the half-diagonal},
\bg\label{Ei}
	E_i \equiv \frac {h_i}{(\new{\prod}_{i=1}^3h_i)^{1/3}} = \frac {s_i}{(\new{\prod}_{i=1}^3s_i)^{1/3}}.
	\old{E_i \equiv \frac {h_i}{(\sum_{i=1}^3h_i)^{1/3}} = \frac {s_i}{(\sum_{i=1}^3s_i)^{1/3}}.}
\eg

Given this definition of the energies $E_i$, $\beta$ is the only free parameter in the Gibbs distribution of Eq.~(\ref{Gib}). This parameter can be fitted to experimental data, for example using a maximum likelihood estimation (MLE). This method consists in maximizing the likelihood function
\bg\label{ML}
	L(\beta)\equiv \prod_{i=1}^6 [p_i(\beta)]^{n_i} = Z(\beta)^{-N} \prod_{i=1}^6 \exp(-\beta E_i n_i),
\eg
where $n_i$ is the number of observations of state $i$ and $N\equiv\sum_{i=1}^6n_i$. In practice, $L(\beta)$ can easily be maximized by minimizing $\ln Z(\beta)+\beta\sum_{i=1}^6E_i f_i$. Different values of $\beta$ can maximize $L(\beta)$ in different experimental conditions. As explained above, small values of $\beta$ are expected in experimental conditions where all six faces appear frequently, while higher values of $\beta$ are expected if the smallest faces appear rarely. Explicitly, in the control experiment I (tossing with a dice cup) smaller values of $\beta$ are expected, than in the control experiment II (free fall from 1~m height). In fact, the MLE yields $\beta=\new{2.69}\old{4.90}$ and $\beta=\new{5.58}\old{10.2}$ for experiments I and II, respectively. The corresponding probabilities of the Gibbs model are displayed in the bottom rows of Table \ref{table_control}. These probabilities (e.g.~43.0\% for faces 3 and 4 in experiment \new{II}\old{I}) lie often between the measured frequencies of the corresponding faces (e.g.\ 43.5\% and 42.5\%), thus suggesting that the model sufficiently describes the data. An explicit $\chi^2$ goodness-of-fit test shows that the predictions of the Gibbs model are indeed statistically consistent with the experimental data. This test will be used again and explained in more detail in the following section.

\section{Two classical experiments revisited}\label{section_classical}

Section \ref{section_gibbs} revealed that the Gibbs model offers a good approximation of the data gathered in the control experiment. The control experiment was based on a single cuboid with three different side-lengths. This section confronts the Gibbs model with other experimental data drawn from the literature. The main purpose of this comparison is to test whether the Gibbs model with a constant parameter $\beta$ can describe a variety of differently shaped cuboids, tossed in similar experimental conditions.

The experiments considered here are summarized in Table \ref{table_experiments}. They were performed by Budden (1980) and Heilbronner (1985), respectively. Both authors used families of $xxy$-cuboids, i.e.~cuboids with equal side-lengths $s_x$ in two orthogonal directions and a different side-length $s_y$ in the third direction. Both authors experimented with a family of $m$ cuboids $j=1,...,m$ ($m=15$ for Budden and $m=7$ for Heilbronner) with identical side-lengths $s_x$ ($s_x=15\rm~mm$ for Budden and $s_x=25\rm~mm$ for Heilbronner), and varying side-lengths $s_{y,j}$. Budden used cuboids ``cut from a mild steel bar whose cross-section was a square of side 15~mm. These were distributed to a class of boys who tossed and rolled them while recording the results.'' By contrast, Heilbronner used cuboids from polyvinylchloride of density  $\approx1,500\rm~kg~m^{-3}$. They were tossed ``in the usual manner, i.e. rolled manually or from a shaker on cloth covered surfaces as well as on linoleum in a ratio of approximately one to one for each set of dice.'' The vast differences in the material and tossing techniques between Budden and Heilbronner suggest that these two datasets are described by different values $\beta$ in the Gibbs model. However, the question to be investigated is whether within each dataset (Budden or Heilbronner) the $m$ cuboids can be described by a constant parameter $\beta$.

\begin{table}[b!]
	\centering
	\begin{tabular}{|c|c|c|c|c|c|c|}
		\hline
		& $s_x$ [mm] & $s_y$ [mm] & $N$ & $n_{xx}$ & $f_{xx}$ [\%] & $p_{xx}$ [\%] \\
		\hline
		\multirow{15}{*}{\rotatebox{90}{\mbox{Budden}}} & 15 & 7.1 & 332 & 304 & 91.6 & 91.0 \\
		& 15 & 9.5 & 840 & 620 & 73.8 & 77.0 \\
		& 15 & 11.2 & 799 & 438 & 54.8 & 63.5 \\
		& 15 & 12.15 & 740 & 367 & 49.6 & 55.4 \\
		& 15 & 13.95 & 516 & 206 & 39.9 & 40.8 \\
		& 15 & 14.5 & 530 & 204 & 38.5 & 36.8 \\
		& 15 & 17.4 & 1011 & 150 & 14.8 & 20.2 \\
		& 15 & 18.45 & 532 & 82 & 15.4 & 16.1 \\
		& 15 & 21.6 & 654 & 34 & 5.2 & 8.1 \\
		& 15 & 23.25 & 606 & 24 & 4.0 & 5.7 \\
		& 15 & 24 & 702 & 12 & 1.7 & 4.8 \\
		& 15 & 25.6 & 609 & 19 & 3.1 & 3.5 \\
		& 15 & 28 & 680 & 6 & 0.9 & 2.1 \\
		& 15 & 31.75 & 275 & 2 & 0.7 & 1.0\\
		& 15 & 39.7 & 503 & 3 & 0.6 & 0.2 \\
		\hline
		\multirow{7}{*}{\rotatebox{90}{\mbox{Heilbronner}}} & 25 & 5 & 2145 & 2089 & 97.4 & 98.4 \\
		& 25 & 10 & 2184 & 1929 & 88.3 & 89.8 \\
		& 25 & 15 & 2103 & 1559 & 74.1 & 72.7 \\
		& 25 & 20 & 2238 & 1244 & 55.6 & 51.7 \\
		& 25 & 30 & 2202 & 421 & 19.1 & 20.5 \\
		& 25 & 35 & 2259 & 239 & 10.6 & 12.4 \\
		& 25 & 40 & 2250 & 162 & 7.2  & 7.6 \\
		\hline
	\end{tabular}
	\caption{Experimental datasets obtained by Budden and Heilbronner using different $xxy$-cuboids. The last column is the prediction of the Gibbs model using $\beta=4.46$ (Budden) and $\beta=3.53$ (Heilbronner).}
	\label{table_experiments}
\end{table}

To model the face-probabilities of a single $xxy$-cuboid, the formalism of Section \ref{section_gibbs} can be simplified, since $xxy$-cuboids only exhibit two macro-states: the $xx$-state, showing one of the two square faces, and the $xy$-state showing one of the four rectangular faces. Given $N$ tosses and $n_{xx}$ observations of the $xx$-state, the corresponding frequency is $f_{xx}=n_{xx}/N$. To model the face-probability $p_{xx}$, the Gibbs model of Eq.~(\ref{Gib}) can be rewritten as
\bg\label{px}
	p_{xx} = Z(\beta)^{-1}\exp(-\beta E_y)
\eg
where $E_x=s_x/(s_xs_xs_y)^{1/3}$, $E_y=s_y/(s_xs_xs_y)^{1/3}$, and $Z(\beta)=2\exp(-\beta E_x)+\exp(-\beta E_y)$. Using Eq.~(\ref{ML}) and $p_{xy}=1-p_{xx}$ the likelihood function for $\beta$ becomes
\bg
	L(\beta) = p_{xx}(\beta)^{n_{xx}}(1-p_{xx}(\beta))^{N-n_{xx}}.
\eg

\begin{figure}[t]
	\centerline{\includegraphics[width=9cm]{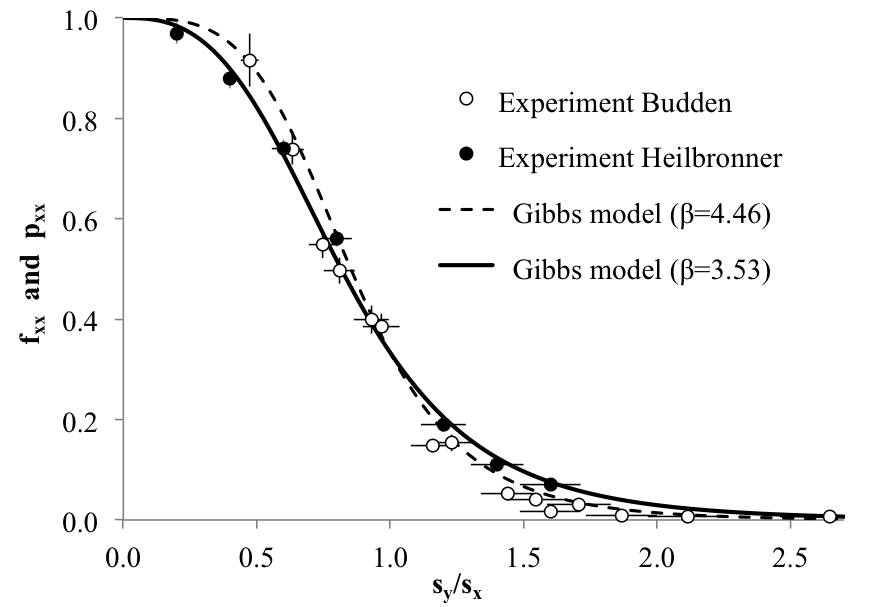}}
	\caption{Measured frequencies $f_{xx}$ and fitted Gibbs probabilities $p_{xx}$ with a constant $\beta$, as a function of the side-ratio $s_y/s_x$ . The experimental values are those listed in Table \ref{table_experiments}. Vertical error bars represent standard deviations of $f_{xx}$ approximated as $\sqrt{f_{xx}/n_{xx}}$, and horizontal error bars are standard deviations associated with 5\% manufacturing errors for the side-lengths.}
	\label{fig_experiments}
\end{figure}

Given a set of $m$ differently sized $xxy$-cuboids $j=1,...,m$ (with respective variables $p_{xx,j}$, $N_j$, $n_{xx,j}$, etc.), the best fitting constant parameter $\beta$ for all cuboids in the same dataset can be obtained by maximizing the global likelihood function
\bg\label{GML}
	\mathcal{L}(\beta) = \prod_{j=1}^m L_j(\beta) = \prod_{j=1}^m p_{xx,j}(\beta)^{n_{xx,j}}(1-p_{xx,j}(\beta))^{N_j-n_{xx,j}}.
\eg
In practice, $\mathcal{L}(\beta)$ can be maximized more easily by minimizing the function $\sum_{j=1}^m[N_j\ln Z_j(\beta)+\beta(n_{xx,j}E_{y,j}+(1-n_{xx,j})E_{x,j})]$. This results in $\beta=4.46$ for the experiments performed by Budden and $\beta=3.53$ for those performed by Heilbronner. The corresponding face-probabilities predicted by the Gibbs model are listed in the last column of Table \ref{table_experiments} and plotted against $s_y/s_x$ in Figure \ref{fig_experiments}. Qualitatively, there seems to be a good agreement between the experimental data and the model.

The rest of this section investigates whether the data is indeed statistically consistent -- in a quantitative way -- with the Gibbs model using a constant $\beta$ per dataset. To do so, a $\chi^2$ test inspired by Gibbons and Chakraborti (2003) is used. The expected number of appearances of the $xx$-state is $N_jp_{xx,j}$ with a variance of $N_jp_{xx,j}$. Therefore, the variance between experimental counts and model-prediction, normalized to the model variance, reads $(N_jp_{xx,j}-n_{xx,j})^2/(N_jp_{xx,j})$. Applying an analogous reasoning to the $xy$-state and summing over all the different cuboids, yields
\bg\label{chi}
	\chi^2 = \sum\limits_{j=1}^m \left[\frac{(N_jp_{xx,j}-n_{xx,j})^2}{N_jp_{xx,j}}+\frac{(N_j(1-p_{xx,j})-(N_j-n_{xx,j}))^2}{N_j(1-p_{xx,j})} \right].
\eg
If $\chi^2/m\leq1$, then the Gibbs model with a constant $\beta$ per dataset fully describes the experimental data; if $\chi^2/m>1$, the experimental data is not sufficiently matched by the model. Explicit calculations yield $\chi^2/m=6.2$ (Budden) and $\chi^2/m=6.6$ (Heilbronner), thus rejecting the hypothesis of a Gibbs model with a constant $\beta$ per dataset at 2.5 standard deviations. In other words, this hypothesis seems to be rejected with a certainty of nearly 99\%. However, this $\chi^2$ test ignores experimental uncertainties of various kinds, which shall now be addressed approximately. 

Potentially, there are various sources of systematic uncertainties in the experimental data. For example, the tossing techniques might have been different for every $xxy$-cuboid. This is particularly plausible in Budden's experiment, where different cuboids were tossed by different children. Further, the cuboids are not perfect due to material and manufacturing errors. To estimate the effect of manufacturing errors on the consistency between the data and the Gibbs model, an extended $\chi^2$ test is performed, which explicitly accounts for uncertainties in the side-lengths $s_x$ and $s_{y,j}$. It is assumed that these side-lengths are only known up to Gaussian errors with standard deviations $\epsilon s_x$ and $\epsilon s_{y,j}$. Hence, $\epsilon$ represents the relative uncertainty of the side-lengths. The hypothesis that the $m$ empirical $f_{xx,j}$ belong to Gibbs distributions with a constant $\beta$ is then tested using a parametric bootstrap test based on 999 independent iterations. In each iteration the following steps are executed.
\begin{itemize}
	\item For every cuboid $j=1,...,m$,
	\begin{itemize}
		\item chose side-lengths $s^\ast_x=G(s_x,\epsilon s_x)$ and $s^\ast_{y,j}=G(s_{y,j},\epsilon s_{y,j})$, where $G(x,\sigma)$ denotes a random number from a normal distribution with mean $x$ and standard deviation $\sigma$,
		\item calculate the probability $p^\ast_{xx,j}$ of the $xx$-state using eq.~(\ref{px}) with the original $\beta$ (4.46 for Budden, 3.53 for Heilbronner) and the new side-lengths $s^\ast_x$ and $s^\ast_{y,j}$,
		\item simulate $N_j$ tossing events, in which the $xx$-state appears with probability $p^\ast_{xx,j}$, and count the number $n^\ast_{xx,j}$,
		\item calculate the corresponding frequencies $f^\ast_{xx,j}=n^\ast_{xx,j}/N_j$.
	\end{itemize}
	\item Use the $m$ values of $f^\ast_{xx,j}$ to estimate the best parameter $\beta^\ast$ via eq.~(\ref{ML}).
	\item Calculate the new probabilities $\tilde{p}_{xx,j}$ using eq.~(\ref{px}) with $\beta^\ast$, $s_x$ and $s_{y,j}$.
	\item Calculate the value $\tilde{\chi}^2$ using eq.~(\ref{chi}) with the probabilities $\tilde{p}_{xx,j}$.
\end{itemize}

\begin{table}[b]
	\centering
	\begin{tabular}{|c|c|c|}
		\hline
		$\epsilon$ & Budden & Heilbronner \\
		\hline
        		0.03 & 0.000 & 0.003 \\
        		0.04 & 0.006 & 0.021 \\
        		0.05 & 0.067 & 0.090 \\
        		0.06 & 0.187 & 0.206 \\
		\hline
	\end{tabular}
    	\caption{$p$-values of $\chi^2$ in the simulated $\chi^2$-distribution to test the hypothesis of the Gibbs model with a constant $\beta$ per dataset (Budden or Heilbronner).}
	\label{table_pvalues}
\end{table}
	
If the original $\chi^2$ is large in comparison to the 999 simulated values of $\tilde{\chi}^2$, the hypothesis of a constant $\beta$ must be rejected.

However, the $p$-values listed in Table \ref{table_pvalues} show that already values around $\epsilon=0.05$, i.e.~manufacturing errors of 5\%, make the experimental data compatible with the hypothesis of a constant $\beta$ for all $m$ cuboids. Measurements of the masses of machine-manufactured wood cuboids similar to those of Budden revealed mass deviations around 7\% between `identical' cuboids, roughly in line with side-length variations of 5\%.

In summary, the Gibbs model with a constant $\beta$ for all the cuboids in a dataset (Budden or Heilbronner) is consistent with the experimental data as long as plausible manufacturing errors are accounted for. Figuratively speaking, the data points in Figure~\ref{fig_experiments} are consistent with the models, as long as the horizontal error bars are included.

\section{Extension to non-cuboidal dice}\label{section_extension}

\begin{figure}[h]
	\centerline{\includegraphics[width=6cm]{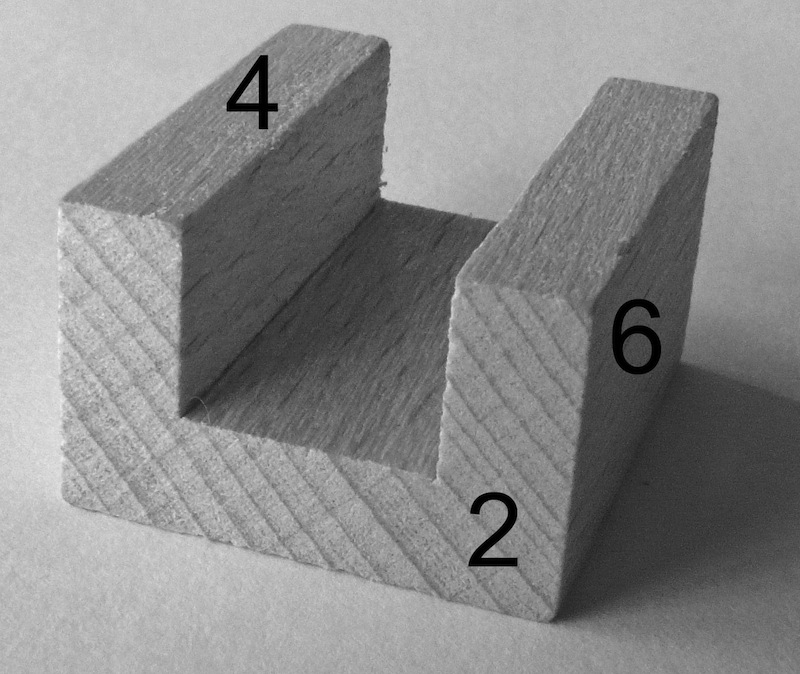}}
	\caption{Image of the U-shaped die. The digits are the indices of the visible faces. Face 1 (opposite face 6) is hidden on the left, face 2 (opposite face 5) is hidden at the back, and face 3 (opposite face 4) is hidden at the bottom.}
	\label{fig_ushape}
\end{figure}

As shown so far, the Gibbs model fully describes the face-probabilities of tossed cuboids within the uncertainties of currently available experimental data. This motivates the idea that the Gibbs model could be extended to more complex dice geometries and inhomogeneous cuboids. A full investigation of this idea lies beyond the scope of this paper, but to provide an illustration the U-shaped die shown in Figure \ref{fig_ushape} is considered. Two experimental runs were performed with this die. In experiment I, the die was tossed $N=1,950$ times onto a hard surface; in experiment II it was dropped $N=150$ times onto a wool carpet. The measured frequencies of the different faces are listed in Table \ref{table_ushape}. Unlike the cuboid, the U-shaped die has no symmetry between the faces 3 and 4. However, the Gibbs model as given in Eq.~(\ref{Gib}) can still be applied using the heights $h_i$ of the center of gravity listed in Table \ref{table_ushape}. To calculate the corresponding energies $E_i$ the $h_i$ are normalized to \new{$h_0=0.5\,(volume)^{1/3}=9.08$~mm, where the volume is now taken as the smallest convex bounding volume}\old{the half-diagonal of 16.45~mm}. The maximum likelihood method yields $\beta=\new{2.77}\old{5.11}$ (experiment I) and $\beta=\new{4.59}\old{8.41}$ (experiment II), respectively. The higher $\beta$ of the second experiment is clearly related to the wool carpet's softness, which tends to destabilize positions with a high center of gravity, thus making the probabilities more skewed towards the most stable positions.

The probabilities of the Gibbs model are consistent with the data in terms of the $\chi^2$ test discussed in Section \ref{section_classical}, hence demonstrating that the Gibbs model extends to non-cuboidal dice.

\begin{table}[t]
	\centering
	\begin{tabular}{|l|c|c|c|c|c|c|}
		\hline
		\bf{Face $i$} & \bf{1} & \bf{2} & \bf{3} & \bf{4} & \bf{5} & \bf{6} \\
		\hline
		Heights of center of gravity $h_i$ [mm] & 10.0 & 11.5 & 7.61 & 5.39 & 11.5 & 10.0 \\
		\hline
		$f_{i}$ experiment I ($N=1,950$) [\%] & 10.6 & 6.9 & 23.9 & 42.5 & 6.8 & 9.3 \\
		$f_{i}$ experiment II ($N=150$) [\%] & 4.7 & 2.0 & 28.0 & 57.3 & 1.3 & 6.7 \\
		\hline
		$p_{i}$ Gibbs model ($\beta=\new{2.77}\old{5.11}$) [\%] & \new{10.6} & \new{6.7} & \new{22.0} & \new{43.4} & \new{6.7} & \new{10.6} \\ % old: 10.4 & 7.3 & 21.9 & 43.6 & 6.5 & 10.4
		$p_{i}$ Gibbs model ($\beta=\new{4.59}\old{8.41}$) [\%] & \new{6.0} & \new{2.8} & \new{20.2} & \new{62.1} & \new{2.8} & \new{6.0} \\ % old: 5.9 & 3.3 & 20.0 & 62.2 & 2.7 & 5.9
		\hline
	\end{tabular}
	\caption{Results of tossing the U-shaped die shown in Figure \ref{fig_ushape}. Note that experiment II has a very small number $N$, thus very large statistical uncertainties on the values $f_i$.}
	\label{table_ushape}
\end{table}

\section{Summary}\label{section_summary}
This paper uncovered that the face-probabilities of a tossed cuboid are well described by the Gibbs model defined via Eqs.~(\ref{Gib}) and (\ref{Ei}). These face-probabilities depend heavily on the tossing conditions -- an effect that can be accounted for by the Gibbs model by adjusting the free parameter $\beta$. Good fits of $\beta$ can be obtained via the maximum likelihood method of Eq.~(\ref{ML}). Typical values of $\beta$ range between \new{1}\old{3} and 10. If differently shaped cuboids are all tossed using similar conditions (material, technique, etc.), then the face-probabilities of all these cuboids can be well approximated using a constant parameter $\beta$, estimated via the global maximum likelihood method of Eq.~(\ref{GML}).

\begin{acknowledgements}
The authors thank Robert Allin for valuable discussions about an earlier version of this paper. D.O.~acknowledges the discussions with Nick Jones. \new{We thank Prof.~Osame Kinouchi for spotting the typo in Eq.~(2) of the original paper.}
\end{acknowledgements}



\end{document}